# Modeling 3D geometry using 1D laser distance measurements with application to cylinder for visualization and evaluating surface quality

Qinwu Xu*
* qinwu.xu2020@gmail.com

**Abstract:** Geometric metrology includes one or two-dimensional (1D/2D) distance or plane measurements, as well as the three-dimensional (3D) scanning. The 1D/2D measuring system is unable to obtain advanced 3D feature, while the 3D scanning system is relatively costly and time-consuming. Accordingly, in this study I developed a 3D geometry and surface prediction method by using 1D laser distance measurements to achieve 3D features while saving cost and time. The model is based on the natural neighbor function for data interpolation and linear model for extrapolation. I implemented the model to the cylinder body for evaluating 3D circularity and surface quality. Results show that the model could achieve reasonable accuracy in constructing the 3D geometry and surface deviation using limited distance measurement (e.g. only 100 data points). It accurately predicts the shape of a curved dent (30.48 mm) and identified minor dents (≤ 0.18mm in depth) which are unable to be detected by eyes and finger touch. It also detects the surface corrugation (1.59 mm) and small local features (6.35 mm) using a measurement resolution of 21.34 mm by 15.49 mm. The predicted 3D circularity is higher than the measured 2D circularity as expected. The 3D model may be extended to other continuous geometry shapes for the future study.

## 1 Introduction

A variety of sensor and metrology technologies including the laser sensor, camera system, and optical devices, have been used to measure the distance and shape of object. The laser sensor is used to measure the distance of points or that in a plane for estimating the circularity of object, which can be called as a 1D or 2D laser gauge measurement. Three-dimensional (3D) scanning system using multiple laser sensors or other optical devices has also been developed and widely used nowadays [López et al. 2020], which can measure the 3D geometry surface of object. For example, the laser scanning system controls the steering of laser beams that are combined with a laser rangefinder to capture multiple surfaces and then form a 3D shape of object [ATOS@ 2020]. Zarubin et al. [2019] used the model-based laser-ultrasonic imaging technique to measure internal geometry of solid parts. Khan et al. [2018] used the laser speckle imaging system for 3D shape in medical application. Schlarp [2018] developed scanning system using the laser line sensors with geometry calibration. Katsuki et al. [2020] used the laser interferometer to measure the sample hole and detect the hole accuracy. Zhang et al. [2016] used the microscopic scattering dark-field imaging method to evaluate surface quality. 3D laser measurement is relatively slow as compared to the 1D/2D measurement. New technologies including structured light have been developed for accelerating 3D measurement [Zhang 2018]. However, the 3D measurement could be also more costly due to the required more components and sensors than 1D/2D measuring system. Laser measurement has also been used for quality control. Fu et al. [2000] used laser technique to measure the radial dimension of ring forgings. Marcio et al. [2019] used laser scanning to evaluate surface quality.

For the 3D scanning systems, mathematical models are used sometimes primarily for measurement improvement or data calibration. For example, Coleman and McCrum [2000] used a mathematical model to quantify accuracy of positioning and reduce the statistical tailor error. Boissonnat and Cazals [2002] used the natural neighbor function to smooth the 3D surface. Chaudhry et al. [2019] used the simulation framework to reduce the noise of 3D laser scanning. Modeling techniques have also been developed to construct 3D geometry based on mathematical models, algorithms, and/or physics for solving specific problems. Jin et al. [2017] reconstructed the 3D human face using the front and side images. Delanoy et al. [2019] combined voxel and image -based predictions to reconstruct 3D sketches. Richter and Roth [2018] developed a Matryoshka Networks model to reconstruct 3D geometry through nested shape layers. William and McCullough [1994] presented the 3D geometry using 2D graphs through mathematical formula of projection. Zhang and Liu [2016] predicted the 3D finishing surface of the part during the machining procedure based on the machining path and physics. Lu et al. [2020] developed a deep learning model for 3D object recognition.

In summary, the 1D/2D laser gauge measurement is fast and costly, but it is unable to obtain advanced 3D features. However, the 3D scanning system is relatively expensive and time consuming. Mathematical models are used for the 3D scanning system primarily for resolution improvement or image calibration as discussed above.

In this study I developed a 3D laser process method to predict the 3D geometry and surface deviation using discrete 1D laser gauge distance measurements. The objective is to develop a method that is as economical as the 1D/2D measurement, while offers advanced 3D features as the 3D scanning system does. I implemented the developed model to a cylindrical body for the evaluation of surface quality including circularity, roughness, and defect such as dent. Based on my literature review, this is the first laser-process method that predicts 3D geometry surface using limited 1D laser distance measurements for the evaluation of geometry and surface quality, which is the main innovation of this paper as compared to existing studies.

## 2  3D Process Method for Geometry Modeling and Surface Quality Evaluation
### 2.1  3D process method

In this paper I developed a 3D process method using limited 1D laser gauge distance measurements. I implemented the model to the cylindrical body and evaluated its 3D circularity and surface defect (e.g., dent and skin corrugation). I also developed a software tool to implement this process method.

In traditional 1D/2D laser gauge measurements on the cylindrical body, the radial deviations (variation to the nominal radius) are measured at discrete points, e.g. 5 heights along the axial direction and 24 angles for each height. The 2D circularity of each height is calculated as the deviation difference (measured max or peak value minus measured min or valley value) as illustrated in Fig. 1. This 1D/2D measurement is fast, but it is unable to quantify 3D features. Users are unable to use these discrete and limit data points to determine the surface properties such as roughness and 3D circularity.

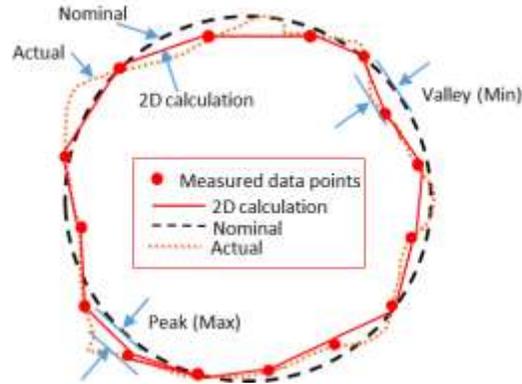

**Fig. 1. Calculation of circularity (peak value minus valley value of measured distances at discrete points).**

In comparison, the developed 3D modeling process method provides a more advanced solution as shown in Fig. 2. The developed process includes three steps: I) laser gauge distance measurements at discrete locations (e.g., only 100 data measurement points on a cylinder body) using a distance laser gauge; II) read the measurement data of deviations; and III) predict the 3D surface and shape using the model and those measured data points as input, and then review the 3D surface and its deviation in both the 3D and unrolled-2D contour views of entire surface. In step I, a sensor sent original wave to the object and a receiver takes the reflected wave to measure distance. Distances to both the sidewall and top of object are measured at discrete points. The color contour of prediction (see Fig. 2) represents the surface deviation value (difference to nominal value).

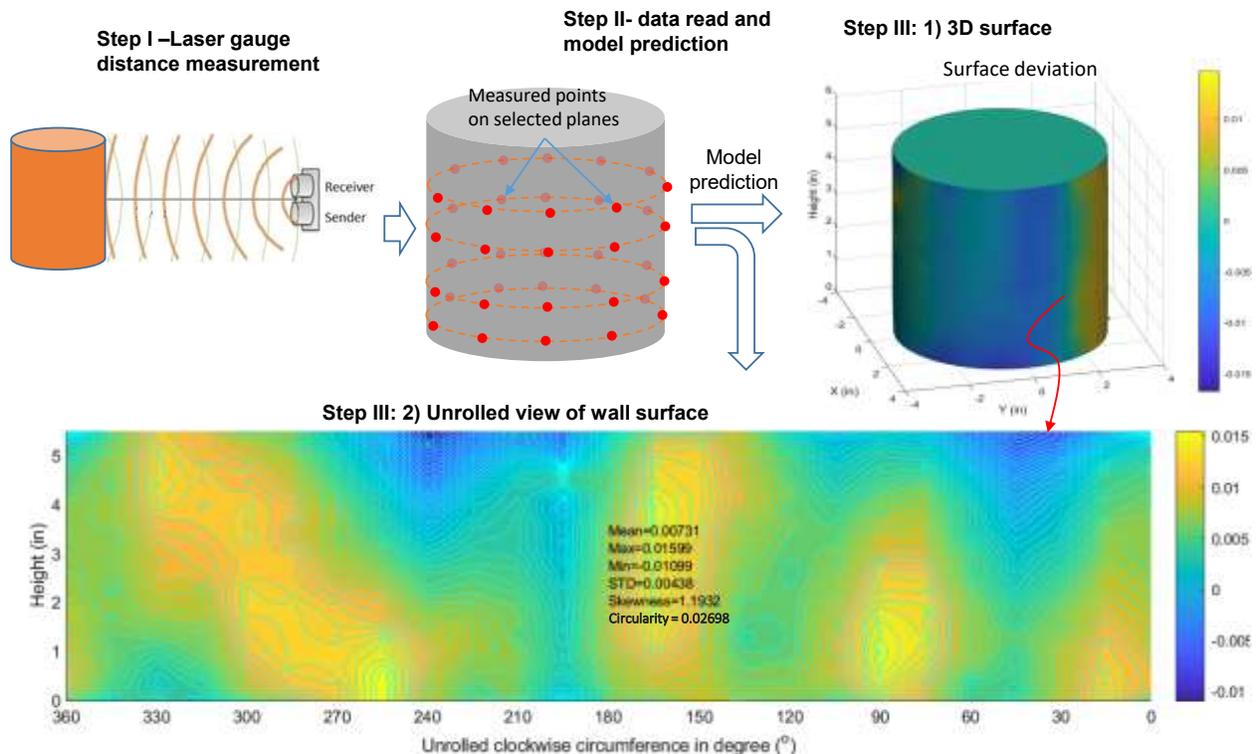

**Fig. 2. 3D process method to predict geometry and surface deviations using laser distance measurements (unit: inch, 1 in= 25.4 mm).**

I developed a geometry model and software tool to predict the 3D surface of the object for the project with multiple parts measured by the laser gauge distance sensor. This modeling process includes the following steps as illustrated in Fig. 3. Firstly, it reads in the data file of all measured parts and extracts the measured deviation data out. It loops over the part ID of the project. Secondly, it predicts the 3D surface for each part using the geometry model as detailed later. Thirdly, it unrolls the predicted 3D geometry surface to a 2D contour display of surface roughness. Lastly, it exports a report of statistics including the mean, max, min, standard deviation, 3D circularity, and skewness of deviation values.

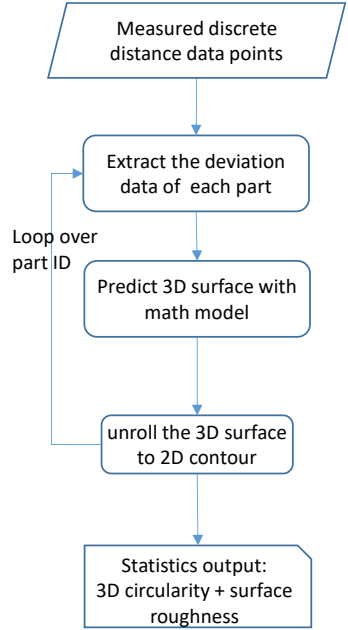

Fig. 3. Flow chart of 3D surface prediction model and software tool.

The 3D circularity is equivalent to the deviation difference between the peak and valley values of the 3D surface. With the predicted 3D surface, I am also able to determine the index of surface roughness which can be quantified by the 3D circularity index, the root mean square $R_{MS}$, and/or the skewness, as follows [Kreyszig 2005], respecively:

$$Circularity = \max(d_i) - \min(d_i), \text{ in } \mathbb{R}^3 \qquad (1)$$

$$R_{MS} = \sqrt{\frac{1}{n}\sum_{i=1}^{n} d_i^2} \qquad (2)$$

$$Skewness = \frac{1}{n}\sum_{i=1}^{n}\left(\frac{d_i-\mu}{\sigma}\right)^4 = \frac{1}{nR_{MS}^3}\sum_{i=1}^{n} d_i^3 \qquad (3)$$

where $d_i$ is the radial deviation of the point $i$, $\mu$ is the mean value, $\sigma$ Is the standard deviation, and $n$ is the total number of data points. Skewness is a statistical measuring of the asymmetry of the probability distribution.

The new development will significantly save cost because the data collection only requires a laser distance sensor. In comparison, the 3D scanning system requires many more hardware components. E.g., the ATOS$^@$ 3D scan device consists of triple scanners, digital microscope, and more [ATOS$^@$ brochure 2020].

## 2.2 3D geometry surface model

I used the geometry models to predict and construct the 3D geometry surface using those limited measurement points (e.g., 24 angles by 5 heights, totaling 100 data points only). The surface prediction includes two steps: 1) interpolate the surface among laser gauge measured distance points, and 2) extrapolate the surface outside of the measured points. I used the linear model for the extrapolation. For the interpolation, I tested the spiral function, cubic spline with $C_2$ continuity, Pchip function with $C_1$ continuity, and natural neighbor function as detailed in the followings. At the end I selected natural neighbor function for the 3D geometry prediction since it provides more accurate results than other models.

### 2.2.1 Spiral function

Spiral function for the 2D geometry shape is defined as follows:

$$x = r(\phi)cos\phi \quad (4)$$
$$y = r(\phi)sin\phi \quad (5)$$

where $r(\phi)$ is radius at angle $\phi$. It becomes a circle when $r(\phi)$ is a constant. Spiral function is continuous and able to simulate the variable radius dynamically. HoIver, the model formula restricts the geometry to a "spiral" shape (may yield to a circle) which is unable to capture the geometry anomaly as shown in Fig. 4A.

### 2.2.2 Cubic spline function with $C_2$ continuity

Spline function has been used to reconstruct 3D surface such including the B-spline function [Yan et al 2016]. To avoid the limitation of spiral function, I then utilized the cubic spline function, as follows:

$$Sp_i(x) = a_i + b_i(x - x_i) + c_i(x - x_i)^2 + d_i(x - x_i)^3 \quad (6)$$

where $i$ is the section number of the curve, $a_i$, $b_i$, $c_i$ and $d_i$ are model parameters, $x$ is the unrolled radial deviation, and $x_i$ is measurement value. I applied the $C_2$ continuity (the second order differential exists and is continuous at the section joint). The model can interpolate the variable geometry shape. However, it produces "overshoots" at section joints (see Fig. 4B).

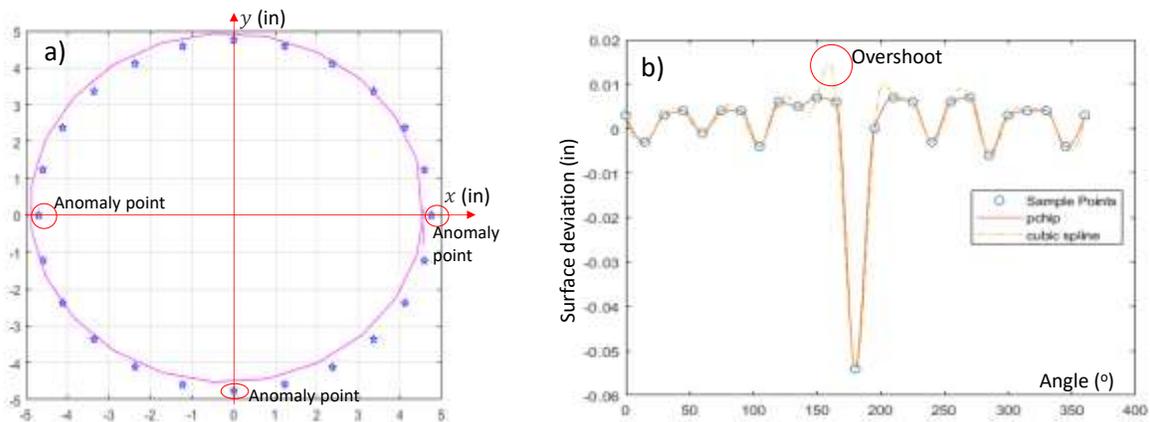

**Fig. 4 Curve fit functions: a) the spiral function is unable to capture anomaly, b) cubic spine function produces "overshoots".**

### 2.2.3 Piecewise Cubic Hermite Polynomial (Pchip) and Makima function with $C_1$ continuity

To address the overshoot issue, the cubic spline function can be modified to the Pchip formula as follows [Kreyszig 2005]:

$$Pchip = a(-x^2 + x^3) + b(3x^2 - 2x^3) + c(x - 2x^2 + x^3) + d(-3x^2 + 2x^3) + d \qquad (7)$$

Where a,b,c, d are model parameters, and $x$ is radial deviation.

The model satisfies the $C_1$ continuity (the first order differential exists and is continuous at the section joint). It eliminates the "overshoots" of cubic spline function as shown in Fig. 4B. However, sometimes the Pchip function may induce relatively sharp connection.

### 2.2.4 Natural neighbor function (NNF)

It is a spatial interpolation method proposed by Sibson [1981]. Following the natural neighbor function, I describe the radial deviation at the object cell as a sum of the deviations of neighbor cells, as follows:

$$d(X) = \sum_1^n w_i(X) d_i(X_i) \qquad (8)$$

Where $w_i(X)$ is the Laplace Iights, $X$ is the interpolated cell centered at the 3D coordinate $(x, y, z)$, $X_i$ is the neighbor cell centered at $(x_i, y_i, z_i)$. The Laplace Iight is expressed as [Belikov et al. 1997]:

$$w_i = \frac{L(X_i)}{d(X_i)} \sum_{m=1}^n \frac{L(X_m)}{S(X_m)} \qquad (9)$$

$L(X_i)$ is the measured interface of the overlapped area betIen the interpolated cell and the neighbor cells, $S(X_i)$ is distance betIen $X_i$ and $X$, and $m$ is the number of overlapped unit cells.

NNF provides a smooth approximation of the object cell that is dependent on its multiple neighboring cells.

Physically, the manufacturing of this Cylinder includes the process of contouring cut under the rotation of a metal bead. During this dynamic and continuous contouring procedure, the shape of a single location point is dependent on its neighbor points as the part is dynamically rotating. Based on this physics, I selected the natural neighbor function that obtains more accurate results than other geometry models discussed above.

## 3 Model Validation

Here I validated the model using two approaches.

### 3.1 Model validation with dense 2D laser gauge measurements

I conducted dense laser gauge distance measurements on a cylindrical part (163.58 mm in diameter by 139.70 mm in height) at 24 angles and 50 heights (total 360 measurement points). Then, I predicted the 3D surface using only 3, 5, 9, and 17 heights of those 50 heights, respectively.

As shown in Fig. 5, the prediction using measurements with 5 or more heights ($\geq$ 21.34 mmx27.94 mm resolution) can capture the distribution overall as compared to that of the 50-height measurement (21.34 mm x 2.79 mm resolution). The prediction using measurements with 9 or more heights ($\geq$21.34 mmx15.49 mm in resolution) achieves satisfied accuracy as it can capture most local features.

The prediction accuracy of local features is dependent on the measurement resolution. E.g., the prediction with 5-height measurement captured one local feature with an edge size of 6.35 mm (see Fig. 5D), but it missed another nearby feature with a size of around 6.86 mm which is predicted by using measurements with 9 or more heights (see Fig. 5C). The prediction accuracy is also dependent on whether the local feature or defect area is shot by one laser gauge point.

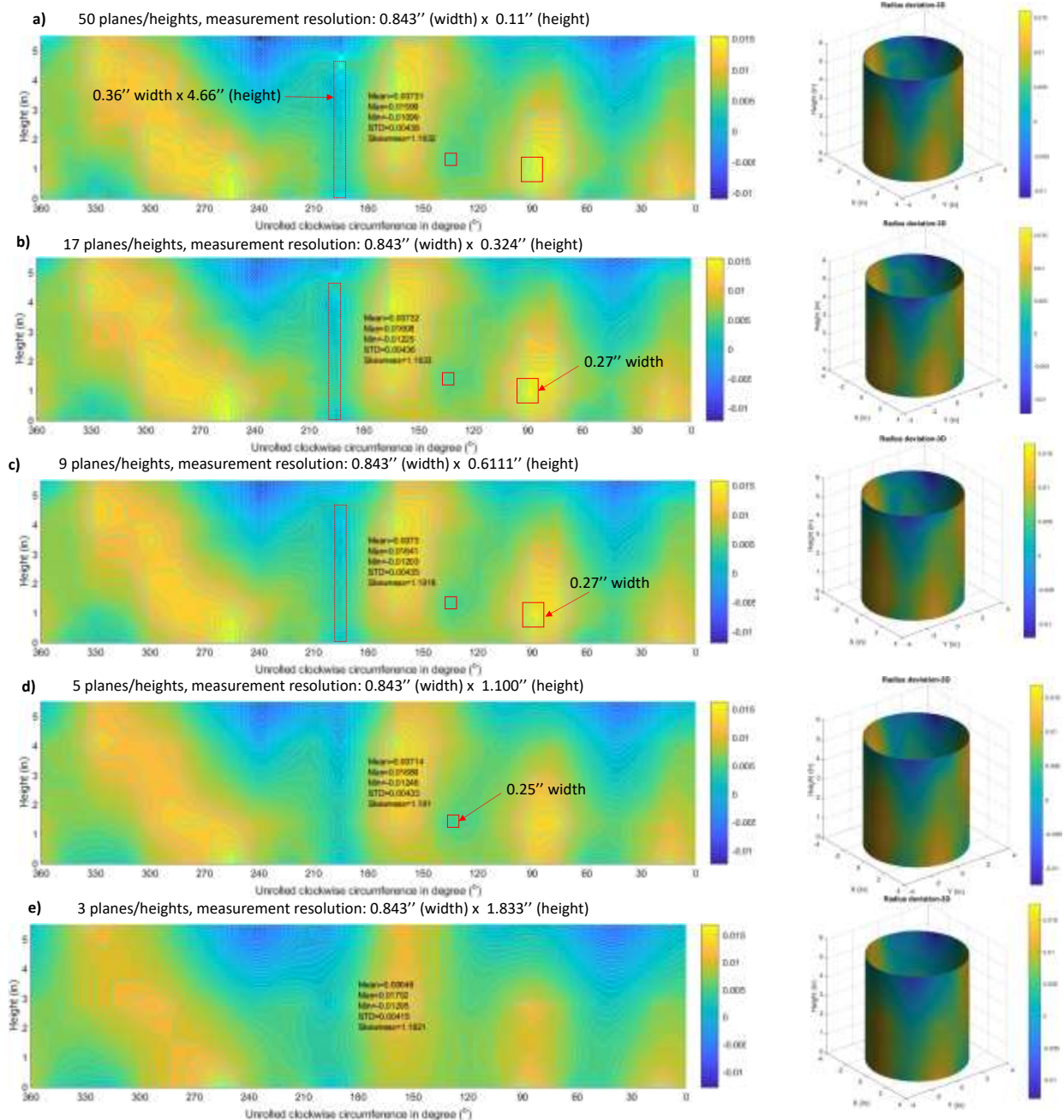

**Fig. 5. Model prediction results of 3D wall surface, using different number of heights of 2D laser gauge measurements (unit: in, 1 in=25.4 mm).**

Fig. 6 plots the summary statistics of the surface property. The prediction using 9-height measurement achieves almost the same mean deviation value as that of the 50-height measurement, while it predicts only 5.4% higher 3D circularity and surface roughness. A 5-height measurement (21.34 mmx27.94 mm resolution) is acceptable for 3D prediction while the ≥ 9 height measurement is recommended (resolution size ≤ 21.34 mmx15.49 mm) to predict local features more accurately.

The measurement time for the 5, 9, 15, and 50 heights is approximately 20, 60, 120, and 300 seconds, respectively. Therefore, using the distance measurements of 5 and 9 heights can save 93% and 80% measurement time of that using the 50-height measurement, respectively. Please note that the 50-heights measurement (resolution of 21.34 mm x 2.79 mm) is still much lower than the actual 3D scanning. This even does not include the saved time for data post-process and the save data storage space. Please also note that the additional time of 3D surface prediction is almost negligible (e.g., 16 seconds for total 50 parts using the parallel computing).

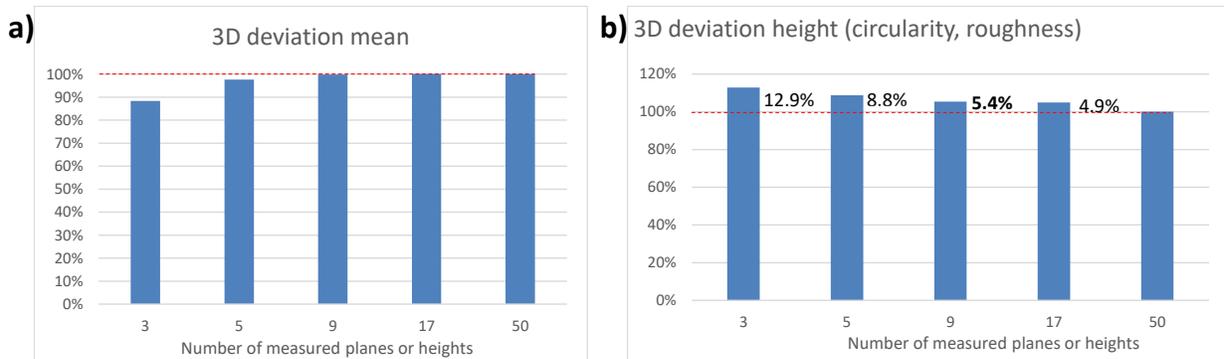

**Fig. 6. Summary of statistics of the 3D surface prediction: a) 3D deviation mean value and b) 3D deviation height (peak value minus valley value) as an index that quantifies 3D circularity and surface roughness.**

### 3.2 Model validation with scanned image

I also used the CT scan image as a second and approximate validation. Fig. 7 plots the predicted 3D surface deviation and shape *vs.* the camera picture of a cylinder (163.58 mm in diameter by 139.70 mm in height).

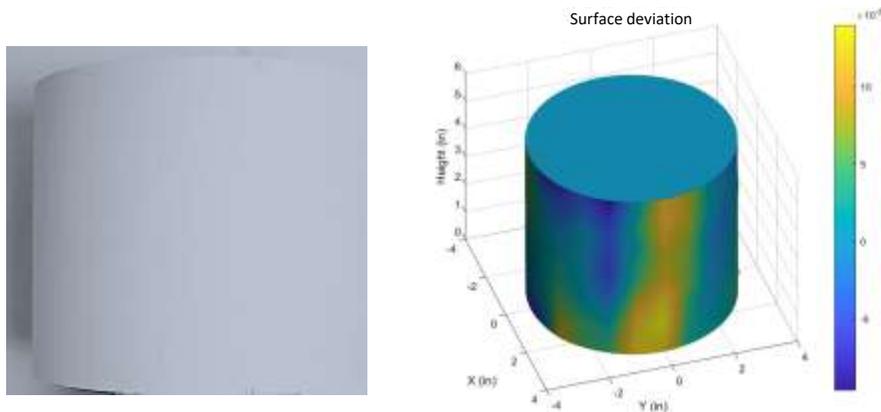

**Fig. 7. Camera picture (left) and 3D prediction of surface deviations including end faces (right, unit: in, 1 in=25.4 mm).**

The scanned 3D surface was unrolled using an approximate method as shown in Fig. 8A. The dark area means that the actual surface falls inside the nominal circle (valley), the white area means that the actual surface falls outside of the nominal circle (peak), and the grey area is located somewhere between the peak and valley. Please note this is only an approximate method, and it is unable to provide specific surface deviation values neither due to the limitation of available facility. In comparison, the laser gauge 3D prediction of deviation values using only 9 measurement heights has shown a consistent trend as that of the scanned one (Fig. 8B *vs.* Fig. 8A). I would not expect the same results exactly due to the approximation method used for the unrolled scan image, as well as the prediction model's limitation. It is difficult for the model to predict the localized small anomaly (e.g., a short crack line) that is missed by laser gauge point.

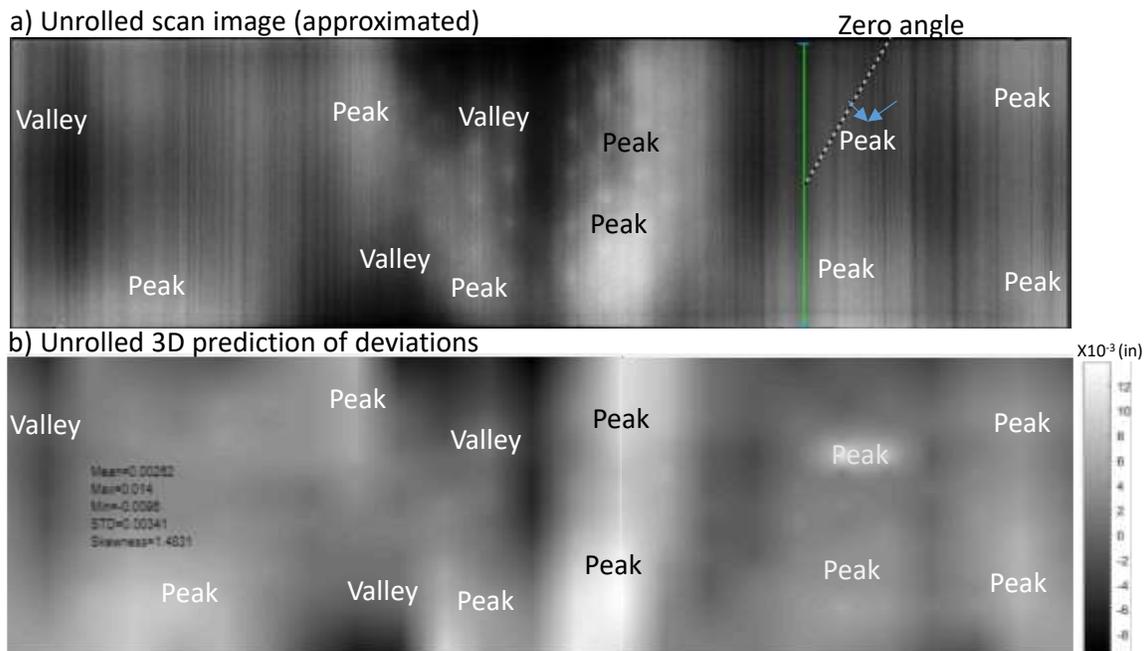

**Fig. 8. Validation of the model: a) unrolled scan image of the wall surface and b) prediction of wall surface deviations, showing consistent trend.**

## 4 Application for Surface Property and Quality Evaluation

### 4.1 Skin dent evaluation

I prepared a cylinder part as shown in Fig. 9 (241.3 mm in diameter by 152.4 mm in height). This part was pressured by robot gripping at multiple locations in process. I then applied the laser gauge 3D process method. Fig. 10 and Fig. 11 show the predicted 3D surfaces. The method identified many skin dents that are undetectable by naked eyes and finger touch (≤ 0.18 mm), other than an externally visible dent with a size of 30.48 mm (width) x 33.02 mm (height) X 1.27 mm (depth).

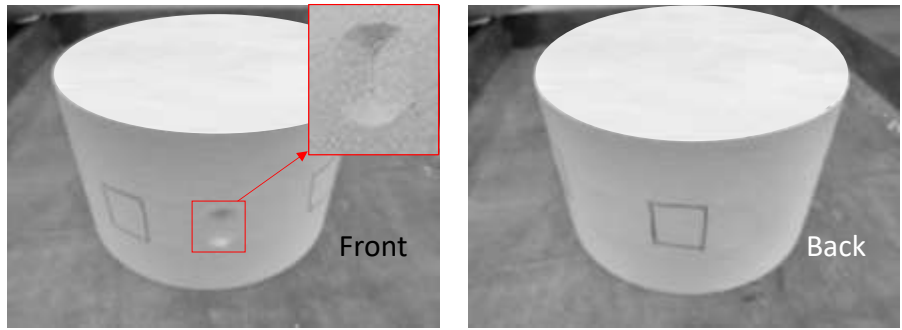

**Fig. 9. Cylinder part has only one visible skin dent, but its rest area looks normal.**

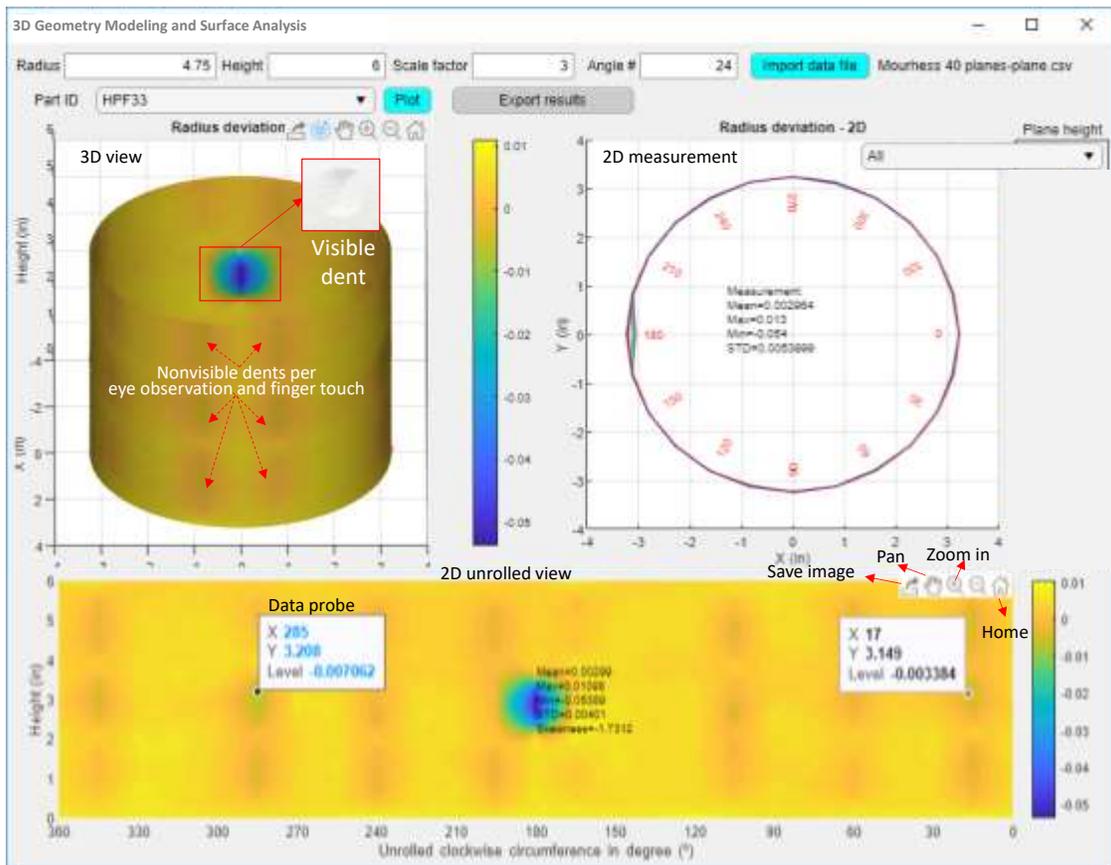

**Fig. 10. Developed software tool and Graphical User Interface (GUI) for the geometry 3D modeling using distance measurements (unit: inch). It identified many minor dents (e.g., ≤ 0.007'' or 0.18 mm in depth) which are undetectable by naked eyes and finger touch.**

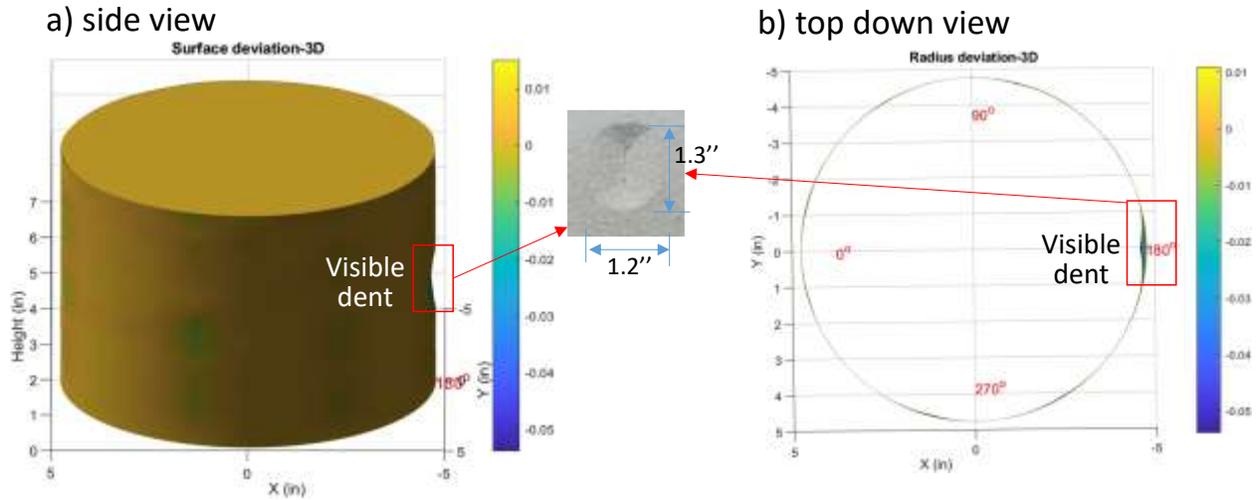

**Fig. 11. The model predicts the 3D shape of the curved-in dent (around 1.2''x1.3'', or 30.48 mm x 33.02 mm) accurately: a) side view and b) top-down view.**

### 4.2 3D circularity and surface roughness

I applied the laser gauge 3D process method to a project lot with 50 cylindrical parts of the same product. I performed laser gauge measurements on 15 heights of each part. Results show that my method could predict some skin corrugation lines (around 1.59 mm, see Fig. 12), as Ill as the variation of surface roughness. In addition, I find that the directly measured 2D circularity under-estimates the predicted 3D circularity as expected (see Fig. 13).

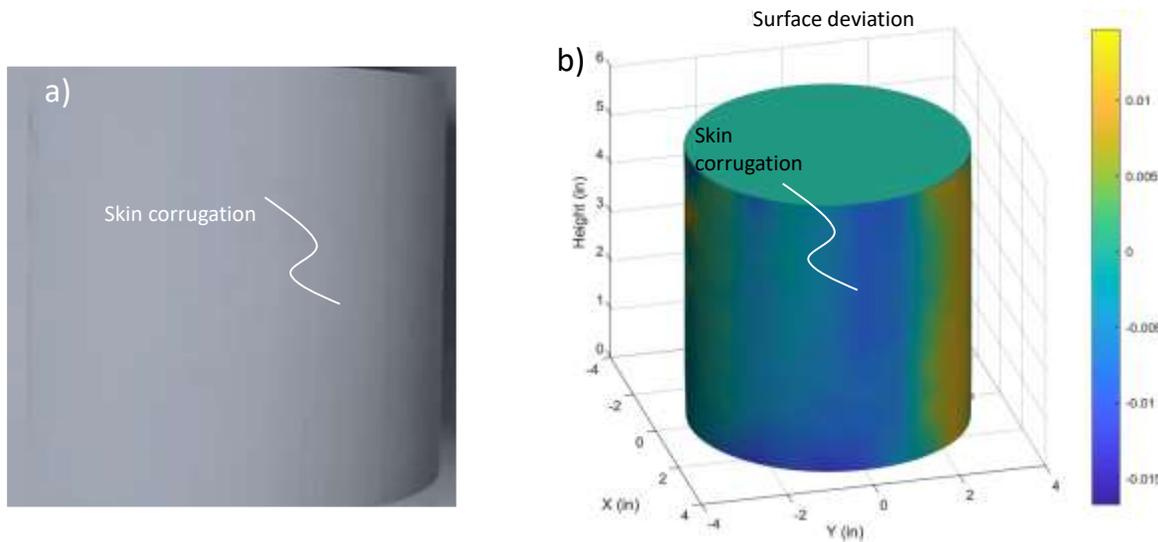

**Fig. 12. Both camera picture and laser gauge 3D surface prediction (unit: in, 1 in=25.4 mm) show the skin corrugation**. *Note: the tiny skin corrugation may not show up on the printed hardcopy.*

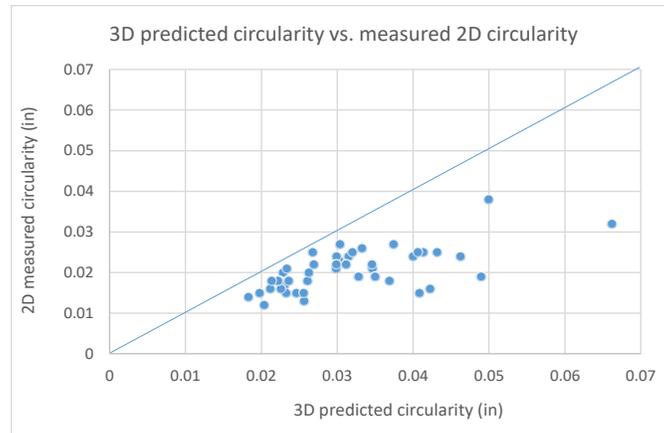

Fig. 13. Measured 2D circularity *vs.* the predicted 3D circularity (1 in = 25.4 mm).

## 5 Conclusion

In this study, I developed a modeling approach to predict the 3D geometry and surface deviation using discrete 1D laser gauge distance measurements. I implemented the method to a cylindrical body for nondestructive evaluation of product surface quality.

The laser gauge 3D process method is computationally efficient. E.g., it takes only 16 seconds to complete data analysis and export report for 50 parts of one project lot. It offers 3D features and quantifies 3D circularity and surface roughness which are lacked by the 1D/2D laser gauge distance measurements. On the other hand, it eliminates some shortcomings of the 3D scanning system that is costly, time consuming, and large in file size. The method accurately predicts the skin dent shape. It also successfully detects some minor defects such as skin corrugations and tiny dent which is unable to be detected by eyes and finger touch.

The model may be extended to other continuous geometry shapes other than cylinder body for the future research.

## 6 Disclosures

The author declares that there are no known competing financial interests or personal relationships that could have appeared to influence the work reported in this paper. This work is only a pure research exploration of the author, which doesn't represent the actual method being used by any organization.